\documentstyle[epsf,onecolumn]{mn}

\title[Weyl focusing effects on image magnification]
{Weyl focusing effects on the image magnification due to randomly
distributed isothermal objects}
\author[Takashi Hamana]
{Takashi Hamana\\
Astronomical Institute, Tohoku University, Sendai 980-77, Japan}
\date{Accepted . Received .}
\pagerange{\pageref{firstpage}--\pageref{lastpage}}
\pubyear{}
\begin{document}
\maketitle
\label{firstpage}

\begin{abstract}
Weyl focusing effects on the image magnifications are investigated 
by using the multiple gravitational lens theory. 
We focus on the gravitational lensing effects due to the small scale
virialized objects, such as galaxies and clusters of galaxies.
We consider a simple model of an inhomogeneous universe.
The matter distribution in the universe is modeled by randomly
distributed isothermal objects.
We found that, for the majority of the random lines of sight, the Weyl
focusing has no significant effect and the image magnification of
a point like source within redshift of 5 is dominated by the Ricci
focusing. 
\end{abstract}


\section{Introduction}
Distance-redshift relation plays an important
role in the astronomy and observational cosmology.
The standard distance, which has been used in the most of previous
studies, is based on a postulate that a distribution of
matter in the universe is homogeneous (e.g., Weinberg 1972).
It has been, however, well recognized that our universe is highly
inhomogeneous on small scales.
Since the inhomogeneities of the mass distribution focus (defocus) the
bundle of light rays (the gravitational lensing effects),
the distance in an inhomogeneous universe
deviates from that in the homogeneous Friedmann universe.
It is, therefore, obvious that a detail understanding of the
propagation of light rays in the inhomogeneous universe is necessary
for correct studies of objects in a distant universe.

Since the pioneering work by Gunn (1967), there has been a lot of
progress in this subject.
Babul \& Lee (1991), among others, studied the effects of the Ricci
focusing by weak inhomogeneities.
They found that the dispersion in image magnifications due
to large scale ($ \ga 0.5h^{-1}$Mpc, where $H_0 = 100 h$km/sec/Mpc)
structures is negligible even for sources at redshift of 4.
They also pointed out that the dispersion is
very sensitive to the nature of the matter distribution on small scales.
The same result was also obtained by Frieman (1996).
He improved Babul \& Lee's study to reflect the recent developments in
numerical and observational studies of the large scale structure.
In these studies, the effects of the Weyl focusing which
induce a shear of light ray bundle were neglected.
Nakamura (1997) examined the effects of the shear on the image magnification
in the cold dark matter model universe with linear density perturbation.
He found that the effect is sufficiently small and
concluded that the Weyl focusing can be safely
neglected for a light ray passing through a linear density
inhomogeneities. 
Jaroszy\'nski et al. (1990) and Wambsganss et al. (1997) used the
multiple gravitational lens theory with a large N-body simulation of
the cold dark matter universe.
Their result agreed well with analytical studies.

The above studies mainly focused on the large scale inhomogeneities, 
whereas the effects of small scale objects, such as galaxies and
clusters of galaxies, have not been fairly taken into account.
Kayser \& Refsdal (1988) investigated the gravitational lensing effects
due to randomly distributed King model galaxies.
They paid a special attention to a high magnification part of the
magnification probability distribution.
Recently Wambsganss, Cen \& Ostriker (1998) studied the gravitational
lensing effects by using the large N-body simulation with an effective
resolution of comoving $10 h^{-1}$kpc.
They, first, shoot the light ray through the lens planes by using the
multiple gravitational lens equation, then
the magnification matrix is determined from the mapping of the light
ray positions between
the image and source plane.
In this procedure, the Ricci and Weyl focusing can not be treated
independently, therefore no discussion is given for the Weyl focusing
effect in their paper.
However, the magnification factor as a function of position in the
source plane and image plane are presented in Figure 4 and 6 of their
paper.
Those Figures show that there is quite a large region that is
demagnified by a small amount, and a few relatively small spots that are
quite highly magnified by the small scale inhomogeneities.

The purpose of this paper is to examine the Weyl focusing effect due
to the small scale inhomogeneities,
and we are not concerned with the effect due to the large
scale structure.
The density distribution in the universe is modeled as a randomly
distributed isothermal lenses. 
The isothermal lens is approximated by virialized objects, such as
galaxy and cluster of galaxies.
This model is similar to the one studied by Kayser \& Refsdal (1988).
However our emphasis is different from theirs, 
i.e. we pay special attention to the gravitational lensing effects on a
majority of light rays.
Although this model is a very simplified and unrealistic one, we believe
that the model is good enough to investigate the essential points of
the Weyl focusing effect due to the small scale inhomogeneities.

This paper is organized as follows: In section 2, we describe our models
and method of simulating the light propagation. 
The results of the simulation are summarized in section 3.
The paper concludes with discussion in section 4.   

\section{Models and Method}
\subsection{Theory of multiple gravitational lensing}
We use the multiple lens equations to trace the propagation of
infinitesimal bundles of light rays.
Schneider, Ehlers \& Falco (1992) deals with the theory of multiple
gravitational lensing in detail. 
Here we simply describe only the aspects which are directly relevant to
this paper.

As was done in previous studies (see e.g., Blandford \& Narayan, 1986
and Kovner 1987), we consider $N$ screens (lens planes)
between an observer $(z=0)$ and source $(z=z_s)$,  
located at redshifts $z_i$ with $i$ runs from 1 to $N+1$ with
$z_{N+1}=z_s$.
In the following, the quantities on the lens and source planes are
described by indices $\{A, B,..\}=\{1,2\}$. 
The position vector of a light ray on the $i$-th lens plane is
denoted by $y_A (z_i)$.
Let $\hat{\alpha}_A (\mbox{\boldmath{$y$}}(z_i))$ denotes the
deflection angle of a light ray at position
$y_A (z_i)$  on the
$i$-th lens plane.
The multiple gravitational lens equation and the evolution equation
of the lensing magnification matrix ${\cal{M}}_{AB}$ are  written as,
\begin{equation}
\label{le}
y_A (z_j) = {{D_j} \over {D_i}} y_A (z_1) - 
\sum_{i=1}^{j-1} D_{ij} \hat{\alpha}_A (\mbox{\boldmath{$y$}}(z_i)),
\end{equation}
\begin{equation}
\label{MAB}
{\cal{M}}_{AB} (\mbox{\boldmath{$y$}}(z_j)) = \delta_{AB} - \sum_{i=1}^{j-1}
{{D_{ij} D_i} \over {D_j}}
\hat{\alpha}_{A,C} (\mbox{\boldmath{$y$}}(z_i)) {\cal{M}}_{CB} (\mbox{\boldmath{$y$}}(z_i)),
\end{equation}
for $2 \le j \le N+1$, 
where $D_{ij}$ $(D_i)$ denotes the standard angular diameter
distance between redshifts of $z_i$ and $z_j$ (0 and $z_i$) with
$i<j$, the comma denotes differentiation with respect to the
components of $y_A(z_i)$ and $\mbox{\boldmath{$y$}}(z_i) \equiv y_A(z_i)$. 
The deflection angle $\hat{\alpha}_A (\mbox{\boldmath{$y$}})$ is
determined by the following equation:
\begin{equation}
\label{2pois}
\hat{\alpha}_A(\mbox{\boldmath{$y$}}) 
= {{4 G} \over {c^2}} \int d^2 \mbox{\boldmath{$y'$}} 
{
{ \mbox{\boldmath{$y$}} - \mbox{\boldmath{$y'$}}}
\over 
{ \left| \mbox{\boldmath{$y$}} - \mbox{\boldmath{$y'$}} \right|^2}
}
\left( \Sigma(\mbox{\boldmath{$y'$}}) -\langle \Sigma \rangle \right),
\end{equation}
where $\Sigma(\mbox{\boldmath{$y$}})$ is the surface mass density and
$\langle \Sigma \rangle$ is its average value.
For a notation convenience, we introduce the following quantities:
\begin{eqnarray}
\label{nal}
\alpha_A (z_i,z_j)
&=& D_{ij} \hat{\alpha}_A (\mbox{\boldmath{$y$}}(z_i)),\\
\label{nt}
{\cal{T}}_{AB} (z_i,z_j)
&=& {{D_{ij} D_i} \over {D_j}}
\hat{\alpha}_{A,B} (\mbox{\boldmath{$y$}}(z_i)).
\end{eqnarray}
The optical tidal matrix ${\cal{T}}_{AB} (z_i,z_j)$ is decomposed into
the Ricci and Weyl focusing terms, respectively:
\begin{eqnarray}
\label{R}
{\cal{R}} (z_i,z_j)
&=& {1\over2} \left( {\cal{T}}_{11}(z_i,z_j) + 
{\cal{T}}_{22}(z_i,z_j) \right), \\
\label{F}
{\cal{F}} (z_i,z_j)
&=&  {1\over2} \left( {\cal{T}}_{11}(z_i,z_j) -
{\cal{T}}_{22}(z_i,z_j) \right)  
+ \mbox{i} {\cal{T}}_{12}(z_i,z_j).
\end{eqnarray}
In general, equation (\ref{MAB}) is not an explicit equation for
${\cal{M}}_{AB}$, since the equation involves a summation over
${\cal{T}}_{AB}$ evaluated on the light ray path, such that one first
has to solve the multiple gravitational lens equation (\ref{le}). 
However for the light rays traveling in regions where $\alpha_A$ and
${\cal{T}}_{AB} < 1$, one can expand
$M_{AB}$ in powers of $\alpha_A$ and ${\cal{T}}_{AB}$
about its value when the light ray is unperturbed. 
We rewrite equation (\ref{le}) as
$y_A(z_j) = y_A^{(0)} (z_j) + y_A^{(1)} (z_j) +
{\cal{O}}(${\boldmath${\alpha}$}${}^2)$
and equation (\ref{MAB}) as 
$
{\cal{M}}_{AB}(${\boldmath${y}$}$ (z_j))=
{\cal{M}}_{AB}^{(0)}(${\boldmath${y}$}${}^{(0)} (z_j)) + 
{\cal{M}}_{AB}^{(1)}(${\boldmath${y}$}${}^{(0)} (z_j)) + 
{\cal{M}}_{AB}^{(2)}(${\boldmath${y}$}${}^{(0)} (z_j))
+...$, 
where $y_A^{(0)}(z_j)$ is the first term of the
right hand side of equation (\ref{le}) and
$y_A^{(1)}(z_j)$ is the second term, but the
deflection angle is evaluated at the unperturbed light ray
position.
Expanding equation (\ref{MAB}) in terms of $\alpha_A$ and
${\cal{T}}_{AB}$,
one finds
\begin{equation}
\label{M0}
{\cal{M}}_{AB}^{(0)} (z_j) = \delta_{AB}, 
\end{equation}
\begin{equation}
\label{M1}
{\cal{M}}_{AB}^{(1)} (z_j) 
=\sum_{i=1}^{j-1} {\cal{T}}_{AB} (z_i,z_j),
\end{equation} 
\begin{equation}
\label{M2}
{\cal{M}}_{AB}^{(2)} (z_j) = 
\sum_{i=2}^{j-1} \sum_{k=1}^{i-1} 
\left[ {\cal{T}}_{AC} (z_i,z_j) {\cal{T}}_{CB} (z_k,z_i)
+ {\cal{T}}_{AB,C} (z_i,z_j) \alpha_C (z_k, z_i) \right],
\end{equation} 
for $3 \le j\le N+1$. 
In the above expressions, ${\cal{T}}_{AB}$ and $\alpha_A$ are
evaluated at the unperturbed light ray position.
The image magnification factor of a point like source is given by 
the inverse of the determinant of the
magnification matrix, i.e. 
$\mu = \left| \det {\cal{M}}_{AB} \right|^{-1}$.
Up to the order of ${\cal{M}}_{AB}^{(2)}$, the
determinant is 
\begin{eqnarray}
\label{recdetM} 
\det {\cal{M}}_{AB} (z_j) &=&  
1 - 2 \sum_{i=1}^{j-1} {\cal{R}} (z_i,z_j) 
+ \left[ \sum_{i=1}^{j-1} {\cal{R}} (z_i,z_j) \right]^2
- \left| \sum_{i=1}^{j-1} {\cal{F}} (z_i,z_j) \right|^2
\nonumber \\
&& + 2 \sum_{i=2}^{j-1} \sum_{k=1}^{i-1}
\left\{
 {\cal{R}}(z_i,z_j) {\cal{R}}(z_k,z_i)
+ \mbox{Re} \right[ {\cal{F}}^{\ast}(z_i,z_j) {\cal{F}}(z_k,z_i) \left] 
+ {\cal{R}}_{,A}(z_i,z_j) \alpha_A (z_k,z_i)
\right\},
\end{eqnarray}
for $3 \le j\le N+1$.
This is our principal equation. 
In general, the effects of Ricci and Weyl focusing
on image magnification are coupled.
However up to this order, they
are not coupled.
We call the terms in the equation (\ref{recdetM}) which involve
Ricci focusing terms as ``Ricci contribution'' and that involve
Weyl focusing terms as ``Weyl contribution''.

\subsection{Truncated singular isothermal sphere lens model}
We adopt the truncated singular isothermal sphere as the lens model.
Its surface mass density is written as
\begin{equation}
\label{2-den}
\Sigma (R) = {{\sigma_v^2} \over {2 G R}} \left( 1+ {R \over {R_G}}
\right)^{-2},
\end{equation}
where $\sigma_v$ is the one-dimensional velocity dispersion and $R_G$
is the half mass radius (Pei, 1993).
The corresponding 3-dimensional mass density runs as $r^{-2}$ $(r \ll
R_G)$ and as $r^{-4}$ $(r \gg R_G)$, and the mass within the radius $R$ is 
\begin{eqnarray*}
m(\le R) = {{\pi \sigma_v^2 R_G} \over G} {R \over {R_G+R}},
\end{eqnarray*}
then the total mass is $m_{tot}=\pi \sigma_v^2 R_G/G$.
This models should provide a fair approximation to virialized
objects, such as a galaxy and cluster of galaxies with isothermal dark
halos.

For the distribution of matter in the universe, we assume that 
the isothermal lenses are randomly distributed with the average mass
density $\rho_L(z)$ and the rest of the matter has the uniform
distribution. 
We also assume that the comoving density of lenses is constant in
time, thus $\rho_L(z)=(1+z)^3 \rho_L(0)$.
Furthermore, we approximate the lensing effects of the isothermal objects,
except the nearest one, to a form of uniform surface mass density.

Under the above assumptions and the circularly symmetric mass
distribution of the lens model, 
the deflection angle is given by
\begin{eqnarray}
\label{al}
\alpha (z_i, z_j)
&=& \sqrt{\alpha_1^2 (z_i, z_j) +\alpha_2^2 (z_i, z_j)} \nonumber \\
&=& 4 \pi \left( {{\sigma_v} \over c} \right)^2 D_{ij} 
\left[ {{R_G} \over {R_G+R}} +{{R_G^2 R} \over {R_c^2(R_G+R_c)}} 
- {{R_G R} \over {R_c^2}} \right]
\nonumber \\
&=& a_{cr} (z_i, z_j) {{D_j} \over {D_i}}
\left[ {{R_G} \over {R_G+R}} -{{R_G R} \over {R_c (R_G+R_c)}}\right],
\end{eqnarray}
with
\begin{equation}
\label{acr}
a_{cr} (z_i, z_j) \equiv 4 \pi \left( {{\sigma_v} \over c} \right)^2
{{D_{ij} D_i} \over {D_j}}, 
\end{equation}
where $R$ is the distance between a light ray position and center of the
nearest lens object in the $i$-th lens plane. $R_c$ is a half of
the mean separation length between the lens objects defined by
\begin{eqnarray}
\label{Rc}
R_c 
&\equiv& \sqrt{{m_{tot}} \over {\pi \Sigma_L^i}} \nonumber \\
&=& \sqrt{a_{cr} (z_i, z_j) R_G } \left[{3\over 2} \Omega_L(0)
(1+z_i)^3 \left( {{H_0} \over {c}} \right)^2 {{D_{ij} D_i}
\over {D_j}} {{c dt} \over {dz}} \delta z \right]^{-{1 \over 2}},
\end{eqnarray}
where $\Sigma_L^i$ is the average surface mass density of the lens
objects in the $i$-th lens plane, and $\delta z$ is the redshift
interval between the
$(i-1)$-th and $i$-th lens planes,
and $\Omega_L(0) \equiv \rho_L(0) / \rho_{cr}(0) = 
\rho_L(0) (8 \pi G /3 H_0^2)$ is the density parameter of lens objects. 
The first term in square bracket of the second line in equation
(\ref{al}) describes the deflection due to the nearest lens object,
the second
term is due to the others, and the last term is due to the background
average mass density of lens objects.
From the deflection angle (\ref{al}), the Ricci and Weyl focusing terms are
immediately given by
\begin{eqnarray}
\label{riccicon}
{\cal{R}}(z_i,z_j) &=& a_{cr} (z_i,z_j) \left[{1\over 2} {{R_G^2} \over {R
(R_G+R)^2}} - {{R_G} \over {R_c (R_G +R_c)}} \right], \\
\label{weylcon}
\left| {\cal{F}}(z_i,z_j) \right| &=& a_{cr} (z_i,z_j) \left[{1\over 2}
{{R_G (R_G+ 2 R)} \over {R
(R_G+R)^2}}\right]. 
\end{eqnarray}

We introduce a compactness parameter $\nu$ as follows;
\begin{eqnarray}
\label{comp}
\nu &=& 4 \pi \left( {{\sigma_v} \over c} \right)^2 {1\over {R_G}} {c
\over {H_0}} \nonumber \\
&\simeq& 1.68 \times \left( {{\sigma_v} \over {200 \mbox{km/sec}}}
\right)^2 \times \left( {{10 \mbox{kpc}} \over {R_G}} \right) h^{-1}.
\end{eqnarray}
This parameter measures the effectiveness of an isothermal object as a
gravitational lens. 
As was pointed out by Kayser \& Refsdal (1988), 
the model with randomly distributed isothermal lens objects is
completely described by two parameters.
The parameters they used depend on the distance between source and
observer. 
On the other hand, we characterize the model by the
compactness parameter $\nu$ and density parameter of lens object
$\Omega_L(0)$ (hereafter we shall simply denote it as $\Omega_L$)
which are independent of redshifts of the source and lenses.

We should note that the optical tidal matrix ${\cal{T}}_{AB}$ becomes larger
than unity at the very central region of the lens object (the inner
region of $ \sim 0.5 \times \mbox{Einstein radius}$),
therefore the validity of assumptions used in deriving the equation  
(\ref{recdetM}) breaks down for light rays passing through that
region.
We now estimate the strong lensing effect on our study in two ways.

The first is based on an order-of-magnitude estimate (section 14 of
Peebles, 1993, and also Futamase \& Sasaki, 1989), we examine a
magnitude of the
tidal matrix ${\cal{T}}_{AB}$.
Suppose the lensing objects are randomly distributed and each with mass  
$M = 2 {\sigma_v}^2 l /G$, 
where $l$ is a characteristic comoving size of a lens
object and is of order $R_G$.
Hence the mean comoving number density of the lens objects is 
$n_L = \Omega_L (3 {H_0}^2 /16 \pi) {\sigma_v}^{-2} l^{-1}$,
so the mean comoving separation distance is
$r_0 = {\Omega_L}^{-1/3} (3 {H_0}^2 /16 \pi)^{-1/3} {\sigma_v}^{2/3}
l^{1/3}$.
Then for a geodesic affine comoving distance of $\lambda$, the light
ray gravitationally encounters such objects
$N_g = \lambda/r_0$
times in average.
At each encounter, the contribution to the tidal matrix is 
$\delta {\cal{T}} = 4\pi ({\sigma_v}^2 /c^2) (r_0/b^2) (\hat{D}_{d}
\hat{D}_{ds}/\hat{D}_s)
\sim 4\pi ({\sigma_v}^2 /c^2) r_0^{-1} (\hat{D}_{d}
\hat{D}_{ds}/\hat{D}_s)$,
where $\hat{D}_{ij}$ is a comoving angular diameter distance with the 
subscript $d$ $(s)$ stands for a lens (source), $b$ is the comoving impact
parameter, and we have assumed that
the mean comoving impact parameter is of order $r_0$.
Since the sign of each contribution will be random, the total
contribution to the optical tidal matrix will be
\begin{eqnarray}
\delta {\cal{T}} \sqrt{N_g} 
&\sim& 
\sqrt{3/4} \sqrt{\Omega_L} \left[ 4\pi
({{\sigma_v} \over c})^2 {1\over {l}} {c \over {H_0}}
\right]^{1\over 2} \langle{{{H_0} \over c} {{\hat{D}_d \hat{D}_{ds}}
\over {\hat{D}_s}}
}\rangle
\left[{{H_0} \over c} \lambda \right]^{1\over 2}\nonumber\\
\label{magoft}
&\sim& \sqrt{\Omega_L} \sqrt{\nu} \langle{{{H_0} \over c} {{\hat{D}_d
\hat{D}_{ds}}
\over {\hat{D}_s}} }\rangle
\left[{{H_0} \over c} \lambda \right]^{1\over 2}.
\end{eqnarray}
The contribution from the direct encounters can be similarly
estimated by noting that the average number of encounters is
$N_d = l^2 \lambda/{r_0}^3 = \lambda \Omega_L (3{H_0}^2/16 \pi)
{\sigma_v}^{-2} l $,
with each encounter contributing
$\delta {\cal{T}}_d = 4\pi ({\sigma_v}^2 /c^2) l^{-1} (\hat{D}_{d}
\hat{D}_{ds}/\hat{D}_s)$ 
with random sign.
The result turns out to be the same as that of gravitational distant
encounters, equation (\ref{magoft}). 
In the case of Einstein-de Sitter universe model, 
the comoving distance $\lambda$ becomes $c/H_0$ at the source redshift 
$z_s = 3$  and the averaged value of distance combination over the
lens redshifts is 
$\langle{({{H_0}/c}) ({{\hat{D}_d \hat{D}_{ds}}/{\hat{D}_s}} })\rangle =
1/6$ for any redshifts of source.
Accordingly, we find that the magnitude of the optical tidal matrix
scales as $\sim 0.2 \sqrt{\Omega_L} \sqrt{\nu}$ for the source redshift
of 3.
In the following we only consider lens models with $\Omega_L \le 1$ and
$\nu \le 1$, therefore a typical value of the optical tidal matrix can 
be expected to be of order $ {\cal{O}}(0.1)$ or lower.
Thus, we can conclude that ${\cal{T}}_{AB}$ is less
than unity for a majority of random lines of sight.

Next we examine the lensing optical depth defined by Turner, Ostriker \& 
Gott, (1984) to quantify the probability of the light rays being
affected by strong lensing.
The differential optical depth for the truncated singular isothermal
sphere model is given by
\begin{equation}
\label{depth}
d\tau = {3\over 2} \Omega_L E (1+z_d)^3 \left({{H_0} \over c}\right)^2 
{{D_d D_{ds}} \over {D_s}} {{cdt} \over {dz_d}} dz_d,
\end{equation}
where
\begin{equation}
\label{E}
E = {1\over 4} {{R_G} \over {a_{cr}}} \left[\left( 1+{4 \over {{{R_G}
\over {a_{cr}}} + {{{R_G}^2} \over {R_c (R_G+R_c)}}}} \right)
^{1\over2}-1 \right]^2. 
\end{equation}
We numerically integrate the last equation for cases of the lens
models with  $(\Omega_L,\nu)=(1,0.1)$, $(1,1)$, $(0.2,0.1)$ and
$(0.2,1)$, and for the Einstein-de Sitter universe model. 
The results are presented in Table 1. 
From the Table 1, it can be found that the probabilities of strong
lensing events are very small except for an extreme model with 
$(\Omega_L,\nu)=(1,1)$.
Even for the extreme model, the probability is not significantly large.
We thus conclude that the strong lensing effects do not
significantly alter our results. 
It can be said from the above two estimations that we can safely use
the perturbative equation (\ref{recdetM}).

\begin{table}
\caption{The lensing optical depth.}
\begin{tabular}{ccccc}
{} & $ (\Omega_L , \nu)=(1,0.1)$ & $(1,1)$ & $(0.2,0.1)$ & $(0.2,1)$ \\
%{} & {} & {} & {} & {} \\
$z_s=1$ & $9.8\times10^{-4}$ & $8.5\times10^{-3}$ & $2.0\times10^{-4}$  
& $1.7\times10^{-3}$ \\ 
$2$ & $2.9\times10^{-3}$ & $2.5\times10^{-2}$ & $5.9\times10^{-4}$ &
$5.0\times10^{-3}$ \\
$3$ & $4.9\times10^{-3}$ & $4.1\times10^{-2}$ & $9.8\times10^{-4}$ &
$8.1\times10^{-3}$ \\
$4$ & $6.6\times10^{-3}$ & $5.5\times10^{-2}$ & $1.3\times10^{-3}$ &
$1.1\times10^{-2}$ \\
$5$ & $8.1\times10^{-3}$ & $6.7\times10^{-2}$ & $1.6\times10^{-3}$ &
$1.3\times10^{-2}$ \\
\end{tabular}
\end{table}

\subsection{Ray shooting}
Since we have assumed the random distribution for lens objects,
the probability of finding lenses in some region on
a lens plane is described by Poisson distribution.
If one sets the redshift interval $\delta z$ to be sufficiently small,
the surface number density of lenses becomes small.
In this case, the lensing effects are mainly due to the nearest lens
and are well approximated by the equations (\ref{al}),
(\ref{riccicon}) and (\ref{weylcon}).
Then the all necessary information
about evaluating the magnification factor (\ref{recdetM}) is obtained
by randomly determining the relative position between a light ray and
the nearest lens object in each lens plane.
We perform Monte-Carlo simulations to trace the propagation of light rays.  
The procedure is described in the following:
\begin{enumerate}
\item 
First of all, we determine the redshift intervals of lens planes to
satisfy the condition that $R_G /R_c$ is sufficiently small.
We set $R_G / R_c < 0.1$. 
\item
The relative positions between a light ray and a center of lens object
are randomly determined in each lens plane.
\item
Summations in equation (\ref{recdetM}) are performed for each
term, and the results are stored in a file.
\end{enumerate}
Steps (ii) and (iii) are repeated for each light ray.

\section{Results}
For the background universe, we only consider the Einstein-de
Sitter universe model, i.e., $\Omega_0=1$ and $\lambda_0=0$. 
For the lens models, we choose the compactness parameter $\nu=1$ and
$0.1$ which roughly correspond to a galaxy scale inhomogeneity and
the scale of a cluster of galaxies respectively.
The density parameter of the lens objects are set to be $\Omega_L=1$
and $0.2$.
We choose the source redshifts $z_s=1$, $2$, $3$, $4$ and $5$.
From the condition $R_G / R_c < 0.1$, the redshift
intervals between lens planes are typically set to be
$\sim 10^{-2}$. 

$10^6$ runs are performed for each model. 
For each run (light ray), we then obtain the Ricci and Weyl
contribution and identically the image magnification factor.
This immediately gives distribution functions of runs on the
Ricci-Weyl contribution plane for all the source redshifts. 
We calculate number densities of results of runs in
Ricci-Weyl contribution plane.
The peaks of the number density and the isodensity contours
which enclose $68\%$ of all runs are presented in Figure 1.
The probability distributions of Ricci and
Weyl contribution for the case of the source redshift of $z_s=3$ in
$(\Omega_L,\nu) = (1,1)$ model are shown in Figure 2.
As is clearly shown in the optical scalar equation (see e.g.,
Schneider et al. 1992), the Weyl contribution is always negative.
Figure 1 reveals that the Weyl contributions in the  majority of light
rays are rather small except in the case where the source redshifts $z_s \ge
3$ in $(\Omega_L,\nu) = (1,1)$ model.
Alternatively, this point can also be shown in Figure 2.
The probability distribution of the Weyl contribution has a narrow
peak centered at very small values, in marked contrast with that
of the Ricci contribution which has a broad distribution. 
Since we have assumed no evolution for lens objects, the
dispersion keeps on spreading in Ricci-Weyl contribution plane even for
high redshift.

%%%%%%%%%%%%%%%% Figure 1 %%%%%%%%%%%%%%%%%%%%%%%%%%%%%%%%%%%
\begin{figure}
\begin{center}
\begin{minipage}{7cm}
\begin{center} 
\epsfxsize=7cm
\epsffile{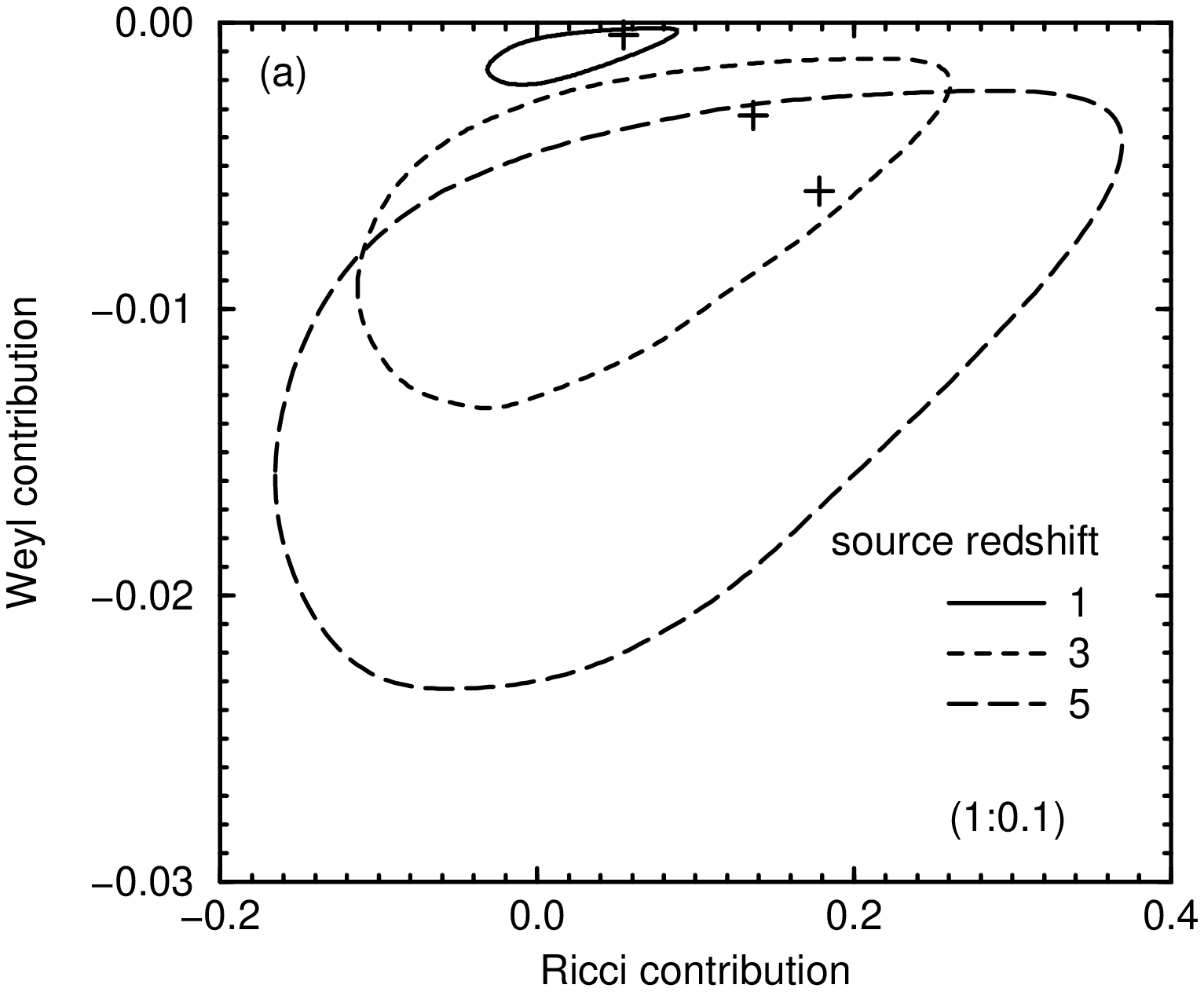}
\end{center}
\end{minipage}
\hspace{0.5cm}
\begin{minipage}{7cm}
\begin{center} 
\epsfxsize=7cm
\epsffile{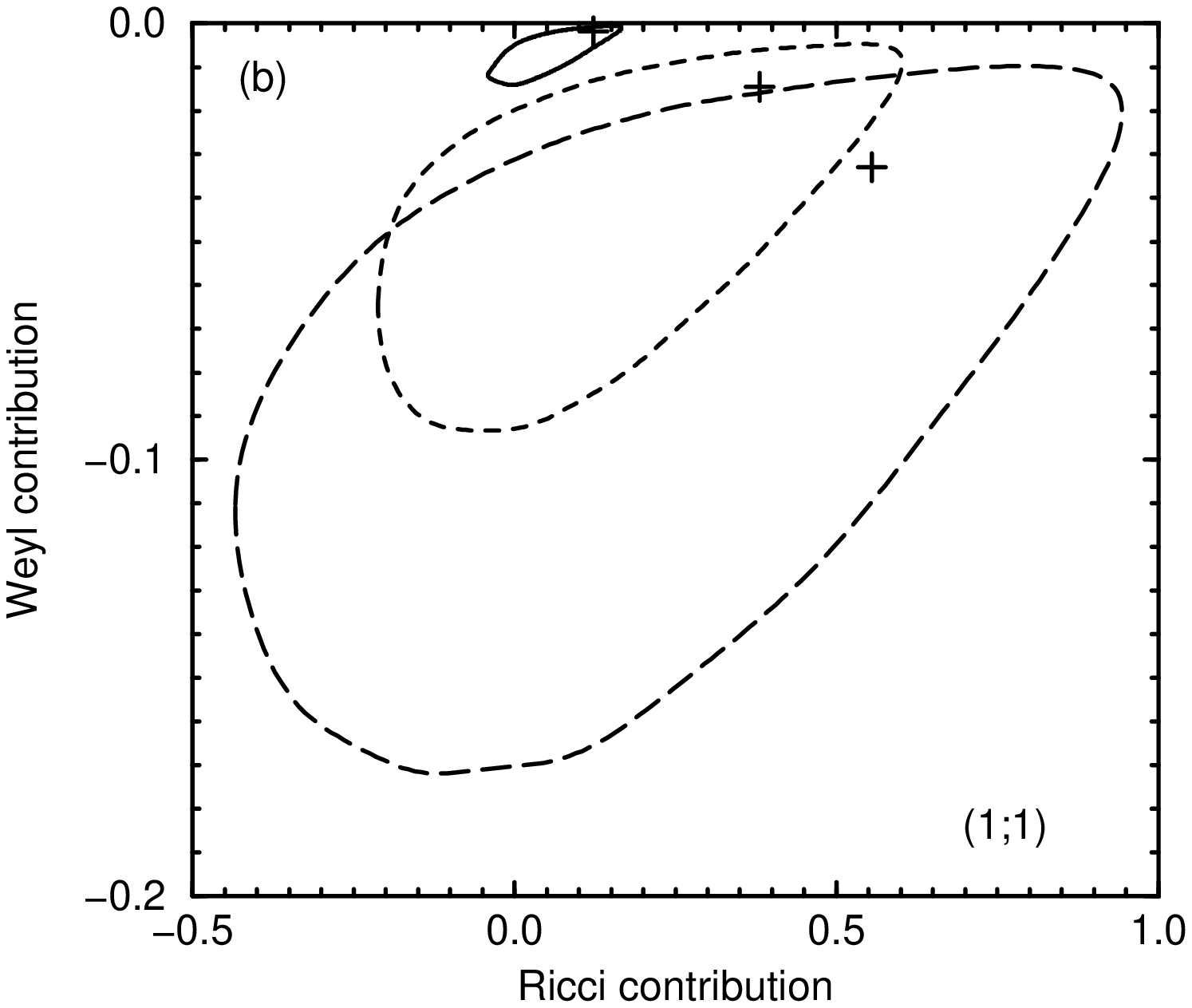}
\end{center}
\end{minipage}
\begin{minipage}{7cm}
\begin{center} 
\epsfxsize=7cm
\epsffile{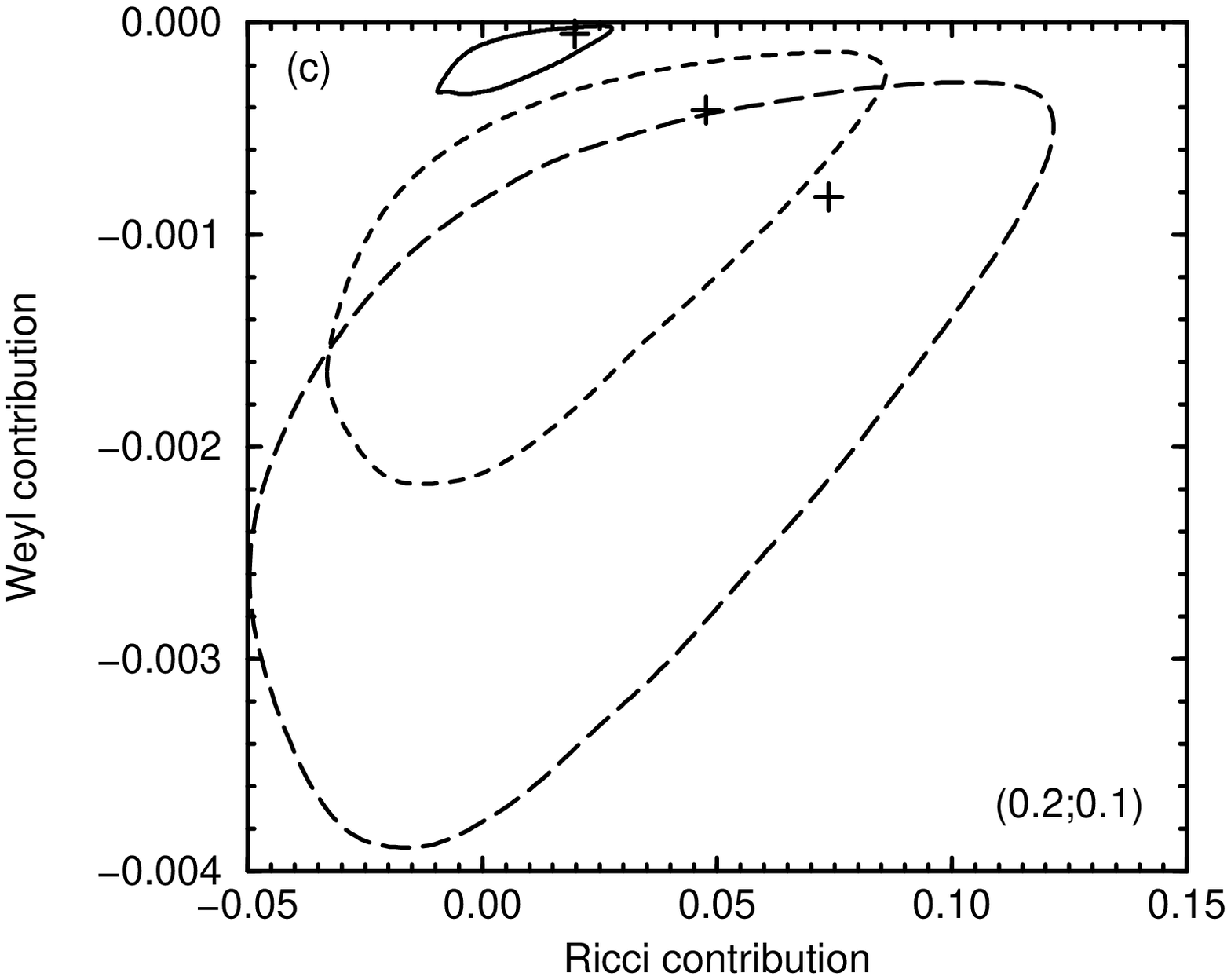}
\end{center}
\end{minipage}
\hspace{0.5cm}
\begin{minipage}{7cm}
\begin{center} 
\epsfxsize=7cm
\epsffile{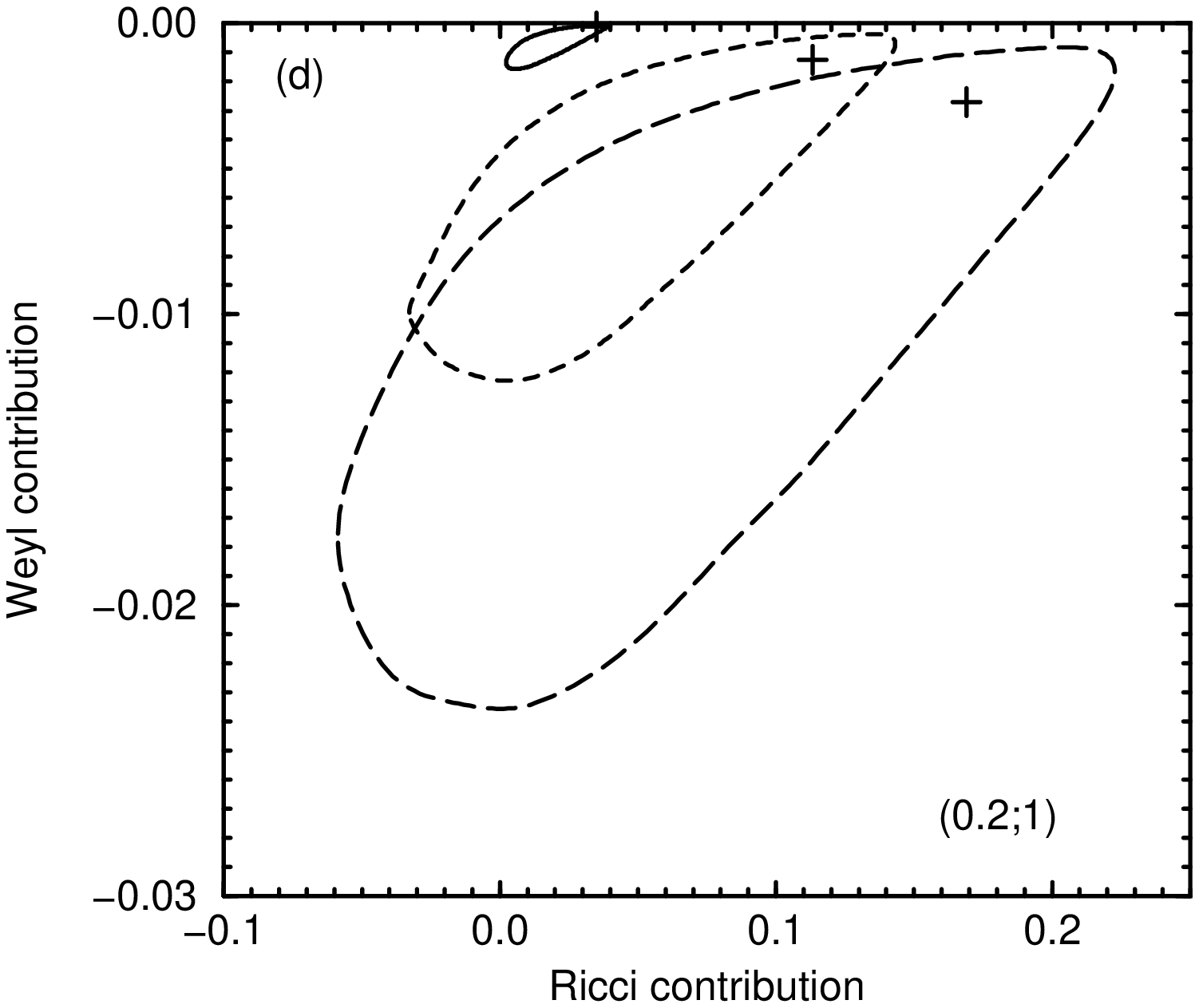}
\end{center}
\end{minipage}
\end{center}
\caption[]{
Isodensity contours which enclose $68\%$ of all runs and
peaks of surface number density of runs.
The peaks are denoted by pluses. The solid lines are for a source
redshift $z_s=1$, dashed lines for $z_s=3$ and long-dashed lines for
$z_s=5$. The lens model parameters $(\Omega_L;
\nu)$ are denoted in each figure.
}
\end{figure}
\begin{figure}
\begin{center}
\begin{minipage}{7cm}
\begin{center} 
\epsfxsize=7cm
\epsffile{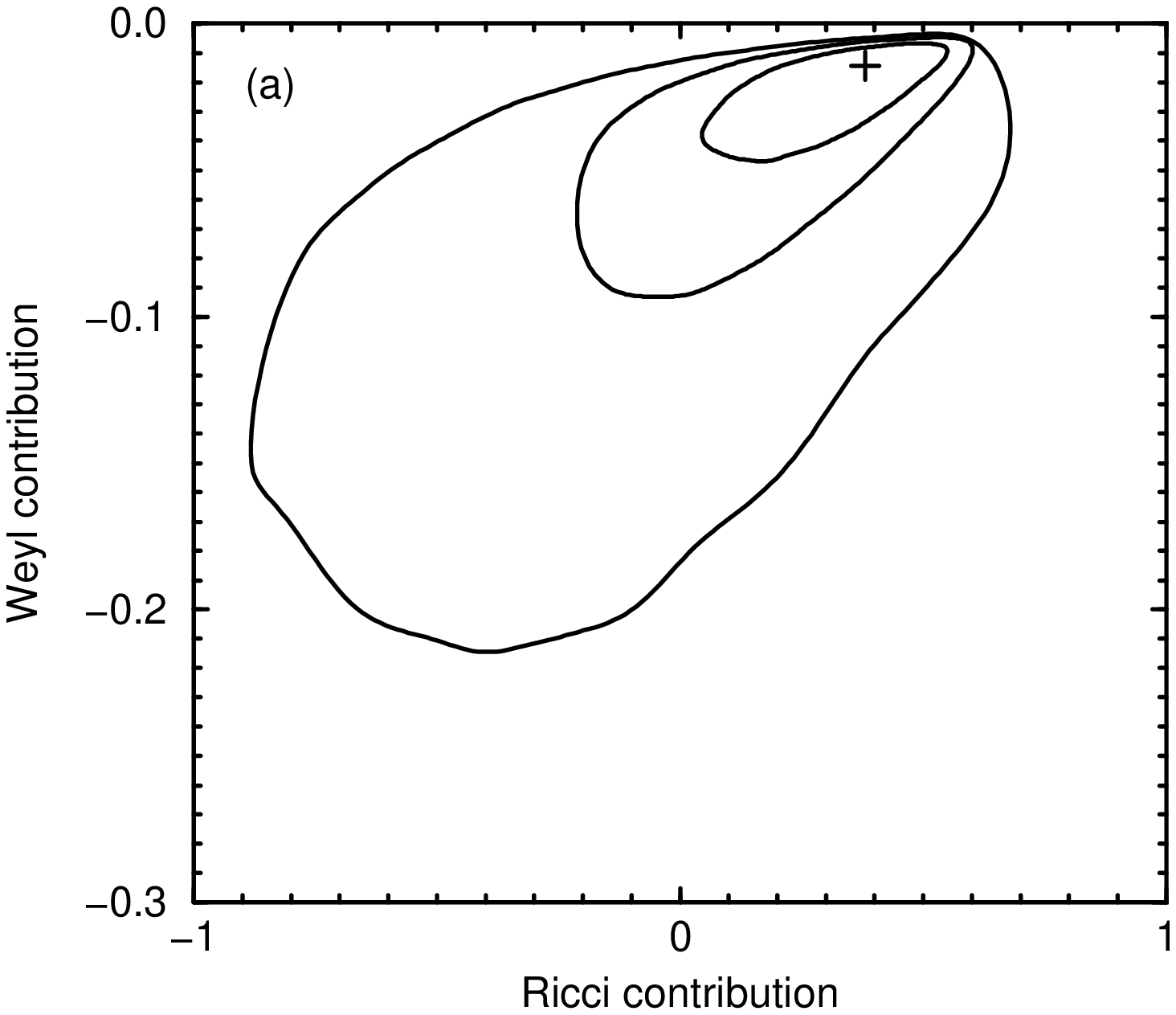}
\end{center}
\end{minipage}
\hspace{0.5cm}
\begin{minipage}{7cm}
\begin{center} 
\epsfxsize=7cm
\epsffile{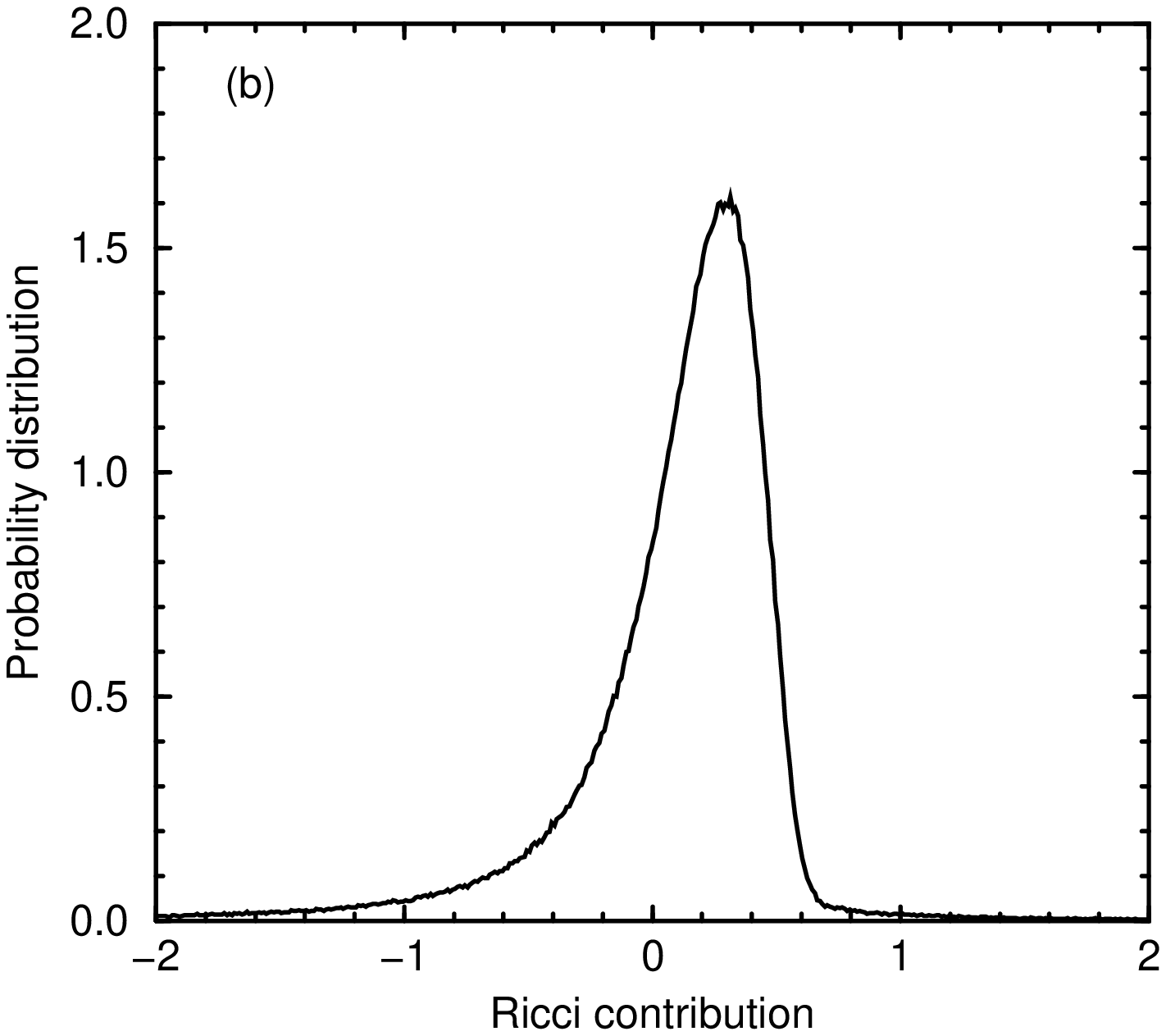}
\end{center}
\end{minipage}
\begin{minipage}{7cm}
\begin{center} 
\epsfxsize=7cm
\epsffile{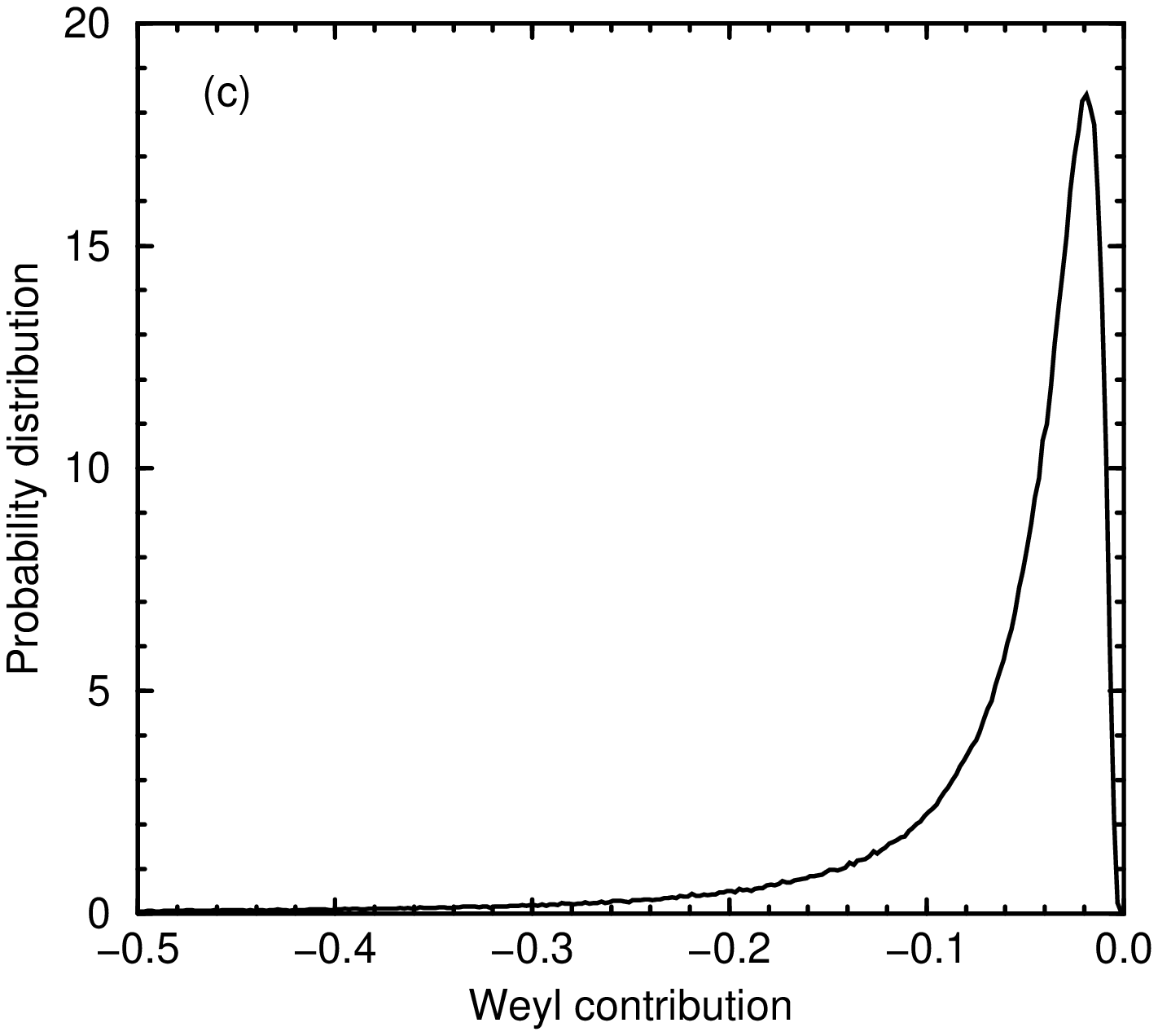}
\end{center}
\end{minipage}
\hspace{0.5cm}
\begin{minipage}{7cm}
\begin{center} 
\caption[]{
Probability distributions for a model with $(\Omega_L;
\nu)=(1;1)$ and $z_s=3$. 
(a) Isodensity contours which enclose $38\%$, $68\%$ and $87\%$ (inner to
outer) of all runs. The plus denotes the peak of the surface number
density of runs. 
(b) The probability distribution of Ricci contributions. 
(c) The probability distribution of Weyl contributions.}
\end{center}
\end{minipage}
\end{center}
\end{figure}

In order to examine the effects of the Weyl focusing on the image
magnification quantitatively, we calculate the magnification factors evaluated
without the Weyl contribution (denoted by $\mu_R$). 
Then we calculate the following quantity which measures the influence of Weyl
contribution on the image magnifications:
\begin{equation}
\label{delmu}
\Delta \mu = \sqrt{{1\over N} \sum_i^N 
\left( 1 - {{\mu_R^i} \over  {\mu^i}} \right)^2 },
\end{equation}
where the summation is taken only over rays with  
$\det {\cal{M}}_{AB} >0$.
The rays with $\det {\cal{M}}_{AB} < 0$ belong to a multiple image
system, therefore the percentage of runs which are excluded in the above
evaluation roughly exhibits the probability of the strong lensing
events among random lines of sight.
At the same time, the above condition also excludes the rays which
pass through the high ${\cal{T}}_{AB}$ region.
The results are presented in Table 2 with the percentages of the
excluded runs (in parentheses).

\begin{table}
\caption{$\Delta \mu$, the effect of the Weyl contribution on the image
magnification with the percentage of runs excluded in this evaluation (in
parentheses).}
\begin{tabular}{ccccc}
{} & $ (\Omega_L , \nu)=(1,0.1)$ & $(1,1)$ & $(0.2,0.1)$ & $(0.2,1)$ \\
%{} & {} & {} & {} & {} \\
$z_s=1$ & $1.8\times 10^{-2}$ (0.16) & $5.7\times 10^{-2}$ (1.0) &
$7.6\times 10^{-3}$  (0.024) & $2.6\times 10^{-2}$  (0.18) \\ 
$2$ & $3.1\times 10^{-2}$ (0.55) & $1.0\times 10^{-1}$ (3.3) &
$1.4\times 10^{-2}$  (0.083) & $4.3\times 10^{-2}$ (0.56) \\
$3$ & $4.0\times 10^{-2}$ (0.96) & $1.3\times 10^{-1}$ (5.6) &
$1.8\times 10^{-2}$  (0.15) & $5.5\times 10^{-2}$ (0.95) \\
$4$ & $4.8\times 10^{-2}$ (1.34) & $1.5\times 10^{-1}$ (7.5) &
$2.1\times 10^{-2}$  (0.21) & $6.5\times 10^{-2}$ (1.3) \\
$5$ & $5.5\times 10^{-2}$ (1.7) & $1.7\times 10^{-1}$ (9.2) &
$2.3\times 10^{-2}$  (0.27) & $7.2\times 10^{-2}$ (1.6) \\
\end{tabular}
\end{table}

Observationally, the strong lensing events among the random lines of
sight are very rare.
For example, the probability of the multiply
imaged quasars in a quasar sample is at most $10^{-2}$ (e.g.,
Claeskens, Jaunsen \& Surdej, 1996).
Combining this fact with our results summarized in Table 2, it may be
reasonably concluded that, as far as our simple matter distribution
model is concerned, the Weyl focusing has a negligible effect on
the majority of light rays even for sources at the redshift of 5.
This result can be naturally explained by the following two reasons:
First, the Weyl focusing is a second order effect on the image
magnification, therefore it becomes important only for the
light rays passing through a very high non-linear (relatively rare)
region. 
Secondly, since we have assumed the random distribution for the
isothermal lenses, the rays coherently affected by the Weyl focusing
are very rare, 
consequently, the majority of light rays are only weakly
influenced by the Weyl focusing.

\section{Discussions}
In this paper, we restricted our study to the gravitational lensing
effects due to the randomly distributed isothermal lenses.
We found that the Weyl focusing effect is small, $\Delta \mu \la 0.1$,
for a majority of light rays.

Lee \& Paczy\'nski (1990) examined the gravitational lensing effects
in randomly distributed clumps with Gaussian surface mass density profile.
They found that the image magnifications are
dominated by Ricci focusing and the Weyl focusing has no significant
effect. 
They only considered a case of a source redshift of $z_s=1.631$, in an
$\Omega_0=1$ and $\lambda_0=0$ universe.
However we have found that,
as far as the random distribution of lens
objects is concerned, the Weyl focusing is rather small even for
higher redshift.

In this study, a correlation of lens objects and large scale structure 
are not taken into account.
The study of the influences of the correlation on the Weyl focusing
lies outside the scope of this paper, and will be examined in future
works.
On the other hand, the Weyl focusing effect due to the large scale
structures is investigated by using N-body simulation (Jaroszy\'nski et
al., 1990) and analytically (Nakamura, 1997).
These two studies show that, although there is an uncertainty of a
normalization of the density power spectrum, the magnitude of Weyl
focusing due to the large scale structure is of the order $10^{-2}
\sim 10^{-1}$ for the source redshifts of $1 < z_s < 5$ (Figure 3 of
Jaroszy\'nski et al., 1990 and Figure 3 of Nakamura, 1997).
Consequently it can be said that, comparing the results of the above
mentioned studies with our results, the Weyl focusing effect due to
the large scale structures is comparable with or larger than that due
to the small scale inhomogeneities.

\section*{Acknowledgments}
The author wishes to thank the referee for valuable comments on the
first version of the manuscript which significantly improved the
quality of this paper.
He would like to thank Professor P. Schneider, Professor T. Futamase,
Dr. M. Hattori and M. Takada for fruitful discussions.
He also would like to thank Dr. P. Premadi for carefully reading and
commenting the manuscript.

\label{lastpage}
\bsp
\end{document}